\documentclass[a4paper,11pt]{article}
\usepackage{jheppub} 
\usepackage{lineno}

\usepackage[utf8]{inputenc} 
\usepackage[T1]{fontenc}    
\usepackage{hyperref}       
\usepackage{url}            
\usepackage{booktabs}       
\usepackage{amsfonts}       
\usepackage{nicefrac}       
\usepackage{microtype}      
\usepackage{lipsum}
\usepackage{fancyhdr}       
\usepackage{graphicx}       
\graphicspath{{media/}}     
\usepackage{placeins}
\usepackage{glossaries}
\usepackage{multirow}
\usepackage{makecell}
\glsdisablehyper

\newacronym{pid}{PID}{Particle identification}
\newacronym{gpt}{GraphPT}{Graph Point Transformer}
\newacronym{tpc}{TPC}{Time Projection Chamber}
\newacronym{gnn}{GNN}{Graph Neural Network}
\newacronym{cepc}{CEPC}{Circular Electron-Positron Collider}
\newacronym{mlp}{MLP}{Multilayer Perceptron}
\newacronym{3d}{3D}{three-dimensional}
\newacronym{2d}{2D}{two-dimensional}
\newacronym{pt}{PT}{Point Transformer}

\newcommand{\dEdx}{d$E$/d$x$}
\newcommand{\dNdx}{d$N$/d$x$}
\newcommand{\gev}{GeV/$c$}
\newcommand{\mum}{$\mu\textrm{m}$}
\newcommand{\mumsq}{$\mu\textrm{m}^2$}

\keywords{Particle Identification, Time Projection Chamber, Cluster Counting, \dNdx, Deep Learning, Point Cloud, Transformer}
\arxivnumber{2510.10628} 

\title{\boldmath \dNdx\ Reconstruction with Deep Learning for High-Granularity TPCs}

\author[a,1]{Guang Zhao,\note{Corresponding author.}}
\author[b]{Yue Chang,}
\author[a,c]{Jinxian Zhang,}
\author[a]{Linghui Wu,}
\author[a]{Huirong Qi,}
\author[a,c]{Xin She,}
\author[a,c]{Mingyi Dong,}
\author[a,c]{Shengsen Sun,}
\author[a]{Jianchun Wang,}
\author[a]{Yifang Wang,}
\author[d]{Chunxu Yu}
\affiliation[a]{Institute of High Energy Physics, Chinese Academy of Sciences, Beijing}
\affiliation[b]{Central China Normal University, Wuhan, Hubei}
\affiliation[c]{University of Chinese Academy of Sciences, Beijing}
\affiliation[d]{Nankai University, Tianjin}

\emailAdd{zhaog@ihep.ac.cn}

\abstract{Particle identification (PID) is essential for future particle physics experiments such as the Circular Electron-Positron Collider (CEPC) and the Future Circular Collider. A high-granularity Time Projection Chamber (TPC) not only provides precise tracking but also enables \dNdx\ measurements for PID. The \dNdx\ method estimates the number of primary ionization electrons, offering significant improvements in PID performance. However, accurate reconstruction remains a major challenge for this approach. In this paper, we introduce a deep learning model, the Graph Point Transformer (GraphPT), for \dNdx\ reconstruction. In our approach, TPC data are represented as point clouds. The network backbone adopts a U-Net architecture built upon graph neural networks, incorporating an attention mechanism for node aggregation specifically optimized for point cloud processing. The proposed GraphPT model surpasses the traditional truncated mean method in PID performance. In particular, for the CEPC baseline TPC, the $K/\pi$ separation power improves by approximately 10\% to 20\% in the momentum interval from 5 to 20~\gev.}

\begin{document}
\maketitle
\flushbottom

\section{Introduction}
\label{sec:intro}
\gls{pid} is a fundamental experimental technique in particle physics. At next-generation facilities such as the \gls{cepc}~\cite{cepc2018cepc, gao2024cepc} and the Future Circular Collider (FCC-ee)~\cite{blondel2022fcc}, hadron identification is expected to remain effective up to momenta of several tens of \gev. Conventional \dEdx\ measurements in current  gaseous detectors~\cite{besiii2009construction, abe2010belle, alme2010alice, anderson2003star}, which determine the total energy loss of a charged particle, cannot meet this requirement due to large fluctuations. Recent advances in electronics and detector technology now enable gaseous detectors to achieve high resolution~\cite{van2025towards, van2026towards, idea2025idea, amaro202350} and to measure individual electron signals, enabling cluster counting (\dNdx). Unlike traditional \dEdx\ methods, the \dNdx\ method directly determines the number of primary ionization clusters~\cite{davidenko1969tnmslation, cataldi1997cluster}, thereby suppressing the impact of secondary ionization as well as fluctuations from energy loss and amplification, and thus offering a significant enhancement in \gls{pid} performance.

The \dNdx\ method can be implemented in a \gls{tpc} when the readout granularity is sufficient to resolve individual electrons. Consequently, the baseline detector design for the \gls{cepc} features a large-volume, high-granularity \gls{tpc} optimized for \dNdx\ measurements~\cite{qi2024feasibility, zhu2023requirement}. Reconstruction, however, presents a significant challenge. In a large-volume TPC such as that proposed for the \gls{cepc}, the detector length is 5.8~m, corresponding to a maximum electron drift distance of 2.9~m. Such a long drift distance results in transverse diffusion on the order of several hundred micrometers, which complicates the use of spatial locality to distinguish overlapping clusters. Furthermore, the \gls{cepc} \gls{tpc} has a pad size of $500 \times 500$~\mumsq\ to balance performance and power consumption. With this pad size and a gas mixture of Ar:CF$_4$:iC$_4$H$_{10}$ (95:3:2), each pad that receives charge from a charged-particle track collects on average about 1.8 primary electrons. Secondary electrons further increase this number, making \dNdx\ reconstruction particularly challenging. 

These challenges make traditional rule-based reconstruction algorithms highly difficult. Deep learning, a statistics-based approach that leverages large training datasets, can extract complex features from the data. In this paper, we propose a \dNdx\ reconstruction algorithm based on deep neural networks, which demonstrates superior performance compared with traditional methods. The remainder of this paper is organized as follows: Section \ref{sec:sim} introduces the detector simulation framework and data samples; section \ref{sec:trad} describes the traditional truncated mean reconstruction method; section \ref{sec:rec} details the deep-learning–based reconstruction algorithm; section \ref{sec:performance} presents the performances; and section \ref{sec:summary} provides the conclusion and plan.

\section{Detector Simulation and Data Samples}
\label{sec:sim}

\subsection{The CEPC TPC}
\label{sec:sub-tpc}
\begin{figure} [htb]
  \centering
  \includegraphics[width=0.5\textwidth]{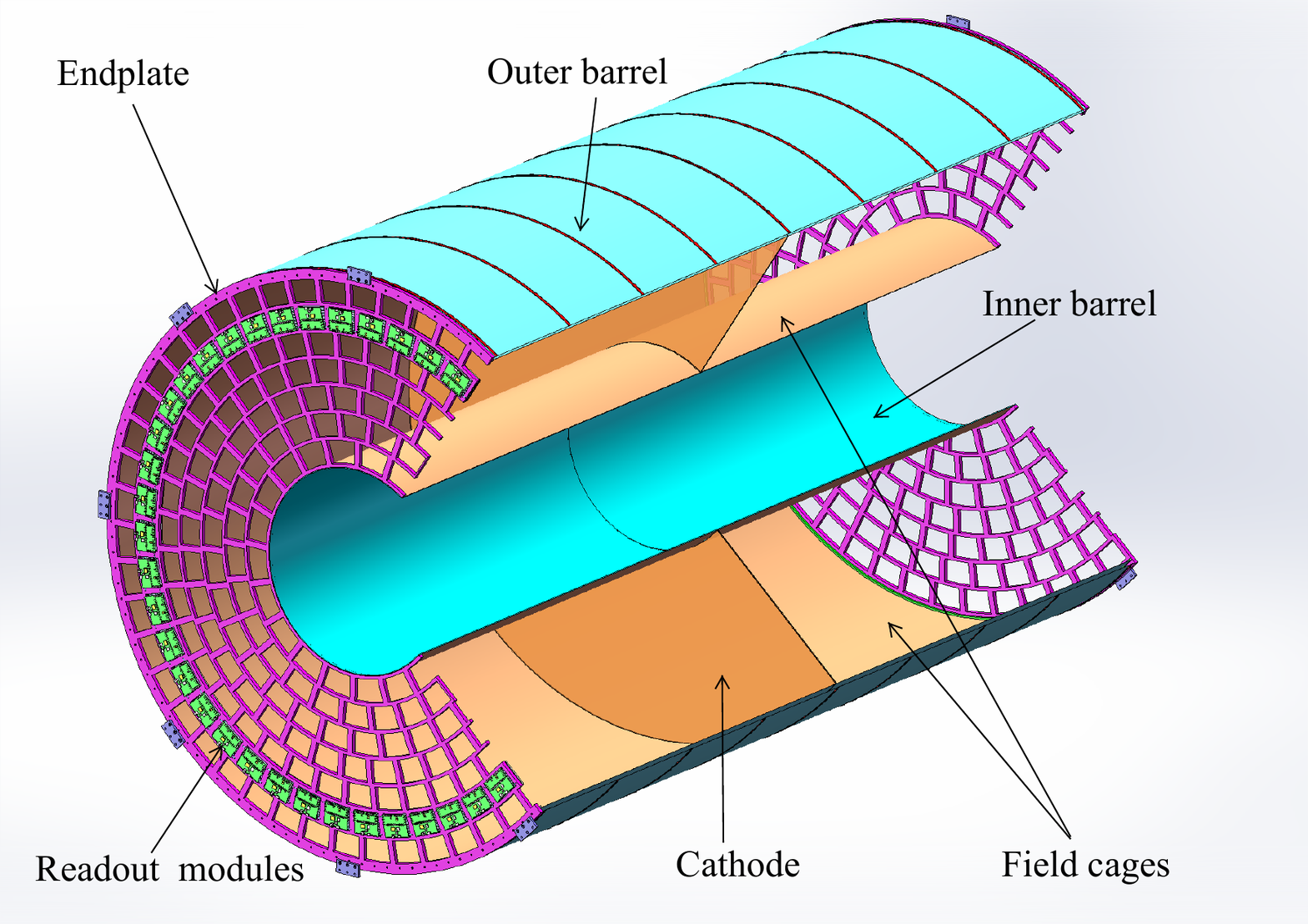}
  \caption{The \gls{cepc} \gls{tpc} is a cylindrical, gas-filled tracking detector whose axis is aligned with the nominal beam direction. The barrel houses a field cage assembly that generates a uniform electric drift field along the Z-axis. Two readout endplates provide mechanical support for the barrel, while a central cathode plane is located at its mid-plane.}
  \label{fig:tpc}
\end{figure}

The \gls{cepc} \gls{tpc} is contained within a cylindrical volume with an outer radius of 1.8~m, an inner radius of 0.6~m, and a total length of 5.8~m (figure~\ref{fig:tpc}). It operates in a 3~T magnetic field. A central cathode, held at a potential of –63~kV, divides the chamber into two drift regions. As charged particles pass through the detector, they ionize the gas mixture, and the liberated electrons drift toward the readout planes. 

Each endcap is equipped with 244 readout modules. Each module consists of a MicroMegas detector~\cite{giomataris1996micromegas}, a readout pad plane and the associated readout electronics. The modules also integrate all necessary high- and low-voltage supply cables, along with a cooling system to dissipate heat generated by the electronics. 

A high-granularity readout~\cite{liu2025development} with a pad size of 500~$\times$~500~\mumsq\ was chosen based on simulation studies, with power consumption also taken into account. Each endcap contains more than 30 million readout channels in total. The fine granularity enables precise \dNdx\ measurements and has the potential to significantly enhance \gls{pid} performance. The key design parameters of the \gls{cepc} \gls{tpc} are summarized in table~\ref{tab:parameters}.

\begin{table}[htbp]
  \centering
  \caption{Key parameters of the \gls{cepc} \gls{tpc}.}
  \label{tab:parameters}
  \begin{tabular}{l|c}
    \hline
    \textbf{Parameter} & \textbf{Value} \\
    \hline
    Radial extension & 0.6--1.8~m \\
    Full length & 5.8~m \\
    Potential at cathode & -63,000~V  \\
    Gas mixture & Ar:CF$_4$:iC$_4$H$_{10}$ = 95:3:2  \\
    Drift velocity & 8~cm/$\mu$s \\
    Transverse diffusion & $30~\mu\textrm{m}/\sqrt{\textrm{cm}}$ \\
    Longitudinal diffusion & $262~\mu\textrm{m}/\sqrt{\textrm{cm}}$ \\
    Gas gain (MicroMegas) & 2000 \\
    Pad size & $500 \times 500$~\mumsq \\
    Total readout modules (per endcap) & 244 \\
    Total readout channels (per endcap) & $\sim$33.5~M\\
    Equivalent Noise Charge (per channel) & $100\,e^-$ \\
    Power consumption & < 100 mW/cm$^2$ \\
    \hline
  \end{tabular}
\end{table}

\subsection{Simulation Framework}
\label{sec:sub-sim}

\begin{figure} [htb]
  \centering
  \includegraphics[width=0.8\textwidth]{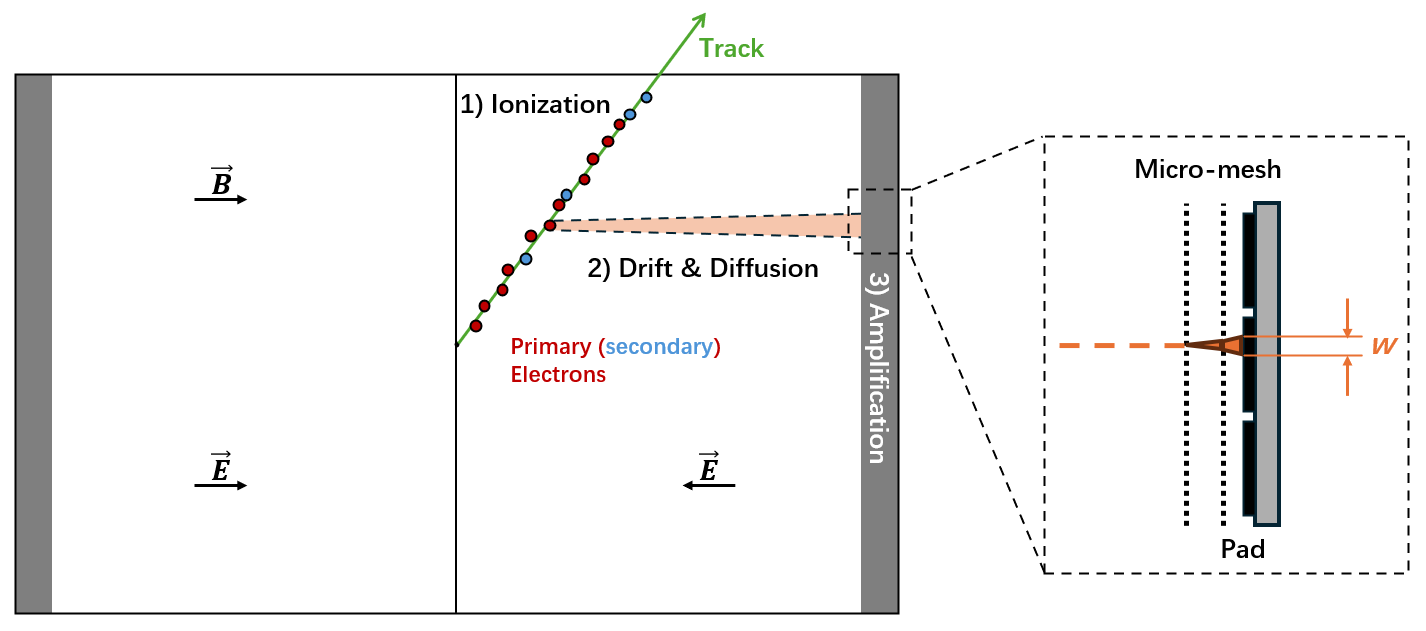}
  \caption{Working principle of a \gls{tpc}. 1) As a charged particle traverses the \gls{tpc} gas mixture, it generates a series of primary and secondary electrons. 2) These electrons drift toward the endcap under the influence of an electric field, undergoing diffusion during their transport. 3) Upon reaching the endcap, the electrons are amplified in the MicroMegas, producing a cluster whose spatial distribution has an effective width $w$.
  }
  \label{fig:tpc_cartoon}
\end{figure}

The \gls{tpc} simulation framework is built on Garfield++~\cite{garfieldpp, veenhof1993garfield} with parameterizations that incorporate detector geometry, ionization processes, electron transport, electron amplification, and signal readout. The latter two stages are informed by results from existing prototype tests~\cite{zhang2020study, chang2025design, liu2025development}. As illustrated in figure~\ref{fig:tpc_cartoon}, a full \gls{tpc} chamber is implemented in the simulation. The gas mixture is set to Ar:CF$_4$:iC$_4$H$_{10}$ (95:3:2). Ionization processes are simulated using the Heed code~\cite{smirnov2005modeling}, which is widely employed for precise ionization modeling in gases.  

\begin{figure} [htb]
  \centering
  \includegraphics[width=1.0\textwidth]{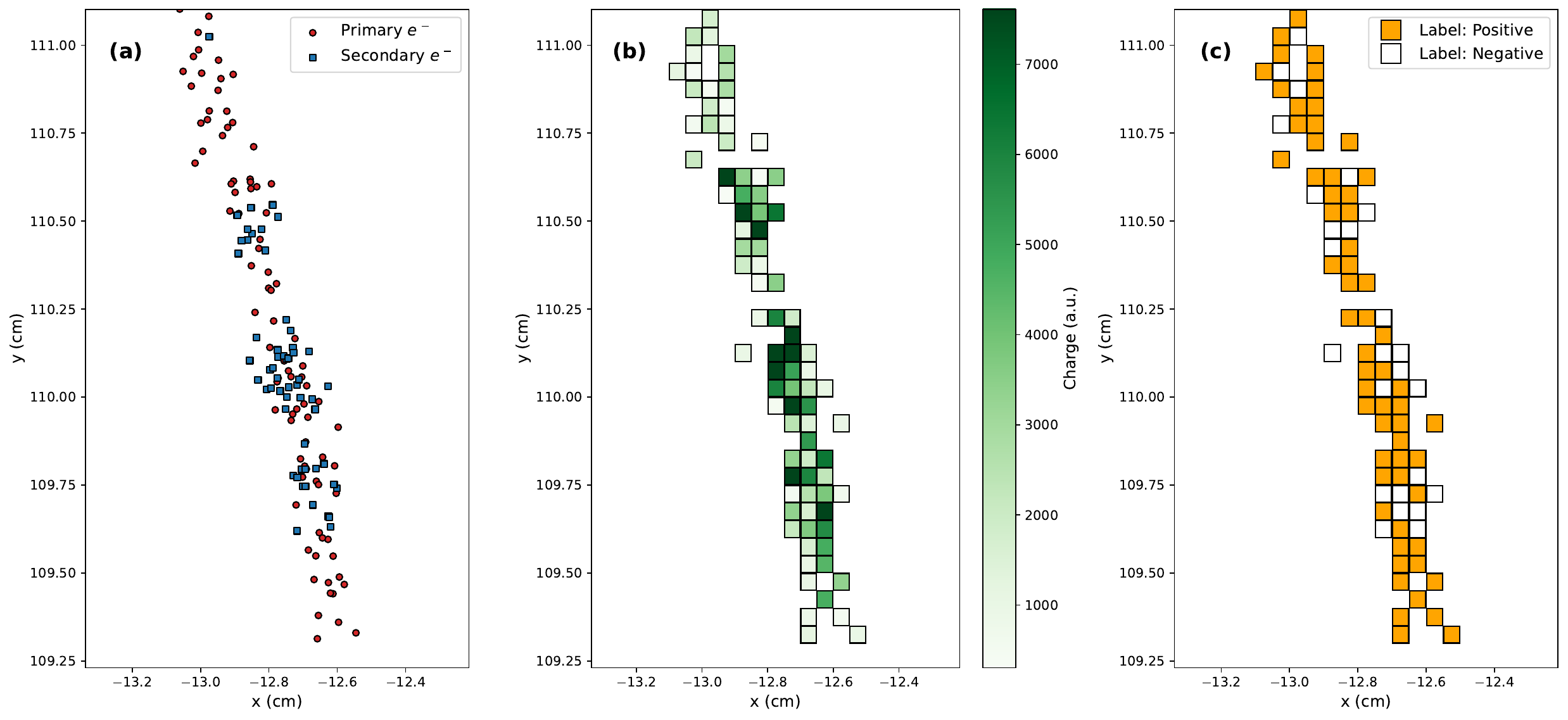}
  \caption{\gls{cepc} \gls{tpc} readout responses for a 2~cm track segment at the endcap. (a) Electrons before amplification, where the red circles represent primary electrons and the blue boxes represent secondary electrons. (b) Energy depositions on the readout pads, with the color bar indicating the deposited charge. (c) Pad labels on the readout pads, where the orange and white colors correspond to labels of positive and negative, respectively.}
  \label{fig:tpc_readout}
\end{figure}

Subsequent processes---electron transport, electron amplification and signal readout---are simulated using parameterized methods, since a full Garfield++ simulation would be computationally prohibitive. The drift velocity and diffusion coefficients of electrons are obtained from the full Garfield++ simulation. The uncertainties in drift position and timing as functions of drift distance are described by a square-root dependence. The final position and timing of each electron after transport are then determined through a sampling procedure. As shown in figure~\ref{fig:tpc_readout}a, the primary and secondary electrons are dispersed around the track trajectory at the readout.

The transported electrons are further amplified in the MicroMegas detector. The avalanche region of the MicroMegas provides a gain of 2000, and the single-electron gain distribution is modeled with a Polya function~\cite{polya1930quelques}. Within the framework, the number of electrons after amplification is sampled according to the Polya distribution, while the spatial density of the amplified electrons is described by a \gls{2d} Gaussian distribution. The effective width $w$ is defined as the the interval spanning $\pm 3\sigma$ of this \gls{2d} Gaussian---corresponding to the width illustrated in the right panel of figure~\ref{fig:tpc_cartoon}---and is approximately 100 $\rm \mu m$.

Finally, the amplified signal is read out according to the pad geometry. Electronic noise is added to all pads using a Gaussian parameterization with a noise level of 100~$e^{-}$ per channel. Figure~\ref{fig:tpc_readout}b presents the final pad responses after accounting for all the processes described above.

\subsection{Data Samples}
\label{sec:sub-sample}
Hadron samples are generated using the simulation framework. For the training dataset, 20,000 pion tracks and 20,000 kaon tracks are produced. These tracks are generated at a fixed polar angle of $60^{\circ}$ with momenta ranging from 0.5 \gev\ to 20.5 \gev. For the testing dataset, hadron tracks are generated at fixed momenta and a fixed polar angle. The chosen momenta are 5 \gev, 10 \gev, 15 \gev, and 20 \gev, all at a polar angle of $60^{\circ}$, with 5,000 tracks generated for each momentum value.

Within the simulation framework, all pad-related hits are recorded for each simulated track. There is a one-to-one correspondence between hits and pads. For each hit, the deposited charge is stored, and its \gls{2d} ($x$, $y$) position is determined from the detector geometry. To obtain the $z$ position, the reconstructed track helix is first projected onto the $x$-$y$ plane, and the on the projected helix that lies closest to the TPC hit is identified as the ``matched'' point. The $z$ difference between the ``matched'' point and the hit is then taken as the hit's $z$ coordinate, which can be converted to timing by dividing by the drift velocity. The charge and timing serve as input features for each hit, while the \gls{3d} hit position ($x$, $y$, $z$) is used to construct the point cloud representation. All input variables are normalized before being fed into the neural network.

In addition to these detector response data, the Monte Carlo (MC) truth information—--essential for neural network training—--is also included. For each pad, the number of primary electrons prior to amplification is stored. As shown in figure~\ref{fig:tpc_readout}c, a pad is labeled positive if it contains at least one primary electron; otherwise, it is labeled negative.

\section{Reconstruction with Truncated Mean Method}
\label{sec:trad}

An efficient reconstruction algorithm is essential for the \dNdx\ method. The ultimate objective is to determine the number of primary electrons---reflecting the ionization characteristics of a charged particle---from the \gls{2d} readout pads. However, as noted in section \ref{sec:intro}, secondary electrons can also generate signals on the readout pads, thereby degrading the \dNdx\ resolution by producing a Landau tail. Consequently, the reconstruction algorithm must be capable of suppressing the influence of these secondary electrons.

A conventional approach for suppressing secondary-electron effects is the truncated mean method, which has been widely applied in \dEdx\ reconstruction~\cite{lippmann2012particle}. In this approach, \dEdx---the energy deposition per unit track length---is used as the reconstruction variable. To mitigate resolution degradation caused by energy-loss fluctuations, typically $N$ independent \dEdx\ measurements $M$ are performed at multiple sampling points along the track within the \gls{tpc} volume. For sufficiently large $N$, the resulting \dEdx\ distribution approaches a Landau distribution. Because of its long tail, the mean value of this distribution is not a reliable estimator of \dEdx. A more robust estimator is the truncated mean, in which the highest measurements---most likely associated with secondary electrons---are discarded. The truncated mean is defined as the average over the lowest $n$ values of the \dEdx\ measurements $M$:
\begin{equation}
    \langle M \rangle_{\alpha}=\frac{1}{n}\sum_{i=1}^{n}M_{i},
\end{equation}
where $M_{i} \leq M_{i+1}$ for $i = 1,...,N-1$ and $\alpha = n/N$ is a fraction. The truncated mean \dEdx\ distribution is well approximated by a Gaussian for sufficiently large $N$ and sufficiently small $\alpha$. For \dNdx, the number of activated pads per unit track length serves as the ionization measurement, since it is approximately proportional to the number of ionization electrons for sufficiently small pads. Consequently, the same truncated mean method can be applied.

\section{Reconstruction with Deep Learning}
\label{sec:rec}

Traditional rule-based approaches, such as the truncated mean method, rely  on predefined selection rules and feature formulations, which limits their ability to address the challenges outlined in section~\ref{sec:intro} for \dNdx\ reconstruction. In contrast, deep-learning–based methods can fully exploit detector information. A deep neural network, trained on large datasets, is capable of capturing hidden correlations among primary and secondary electrons as well as noise contributions. Therefore, we propose a deep-learning method, termed \gls{gpt}, for \dNdx\ reconstruction.


\subsection{Related Work}
\label{sec:sub-relate}

The transition toward modern machine-learning in high-energy physics is driven by the need to resolve complex, non-linear patterns within the high-granularity data produced by next-generation detectors. A key breakthrough in this area is the self-attention mechanism~\cite{vaswani2017attention}, which allows the algorithm to simultaneously evaluate the relationship between every individual signal ``hit'' and every other hit across the detector volume. This capability is crucial for capturing the global information of a particle event more effectively than local, rule-based algorithms~\cite{qu2022particle, ratnikov2021fast, van2025transformers}.

To handle the spatial distribution of these hits, point-based architectures such as PointNet++~\cite{qi2017pointnet, qi2017pointnet++} introduce a hierarchical representation that enables the model to analyze fine-scale hit patterns while simultaneously learning broader track-level information. This approach can be further enhanced by the \gls{pt}~\cite{zhao2021point, wu2022point, wu2024point}, which incorporates self-attention to dynamically weight neighboring hits based on their physical characteristics and relative spatial distances. In parallel, \glspl{gnn}~\cite{scarselli2008graph} treat detector data as a graph of interconnected nodes, where edges encode physical proximity. By learning over this graph structure, \glspl{gnn} can effectively capture geometric shapes, neighborhood symmetries, and any other rational patterns intrinsic to particle interactions~\cite{duarte2022graph, qu2020jet}.

These advancements are particularly relevant for \dNdx\ reconstruction. Beyond traditional rule-based approaches~\cite{caron2014improved, caputo2023particle, aoki2022double}, several pioneering studies have applied neural networks to \dNdx\ reconstruction, particularly in drift chambers, where signals are recorded as one-dimensional waveforms. Long Short-Term Memory networks and Dynamic Graph Convolutional Neural Networks (DGCNNs)~\cite{wang2019dynamic} have been applied to peak detection and clustering tasks with simulated data, achieving performance superior to traditional derivative-based and timing-based methods~\cite{tian2025cluster}. Furthermore, an innovative domain adaptation approach~\cite{zhao2024peak}, leveraging optimal transport theory~\cite{kantorovich1942translocation}, has been developed and successfully applied to test beam data.

\subsection{Design Principle}
\label{sec:sub-design}
For the \gls{tpc}, each hit provides a \gls{3d} spatial coordinate. Consequently, a \gls{tpc} track can be naturally represented as a point cloud. Inspired by the \gls{pt}, a U-Net-based hierarchical architecture is adopted as the backbone. In this work, we further extend the framework by representing the irregularly distributed hits in \gls{3d} space as graphs, where the attention mechanism from the original \gls{pt} is incorporated into the node aggregation process. This integration of \glspl{gnn} allows more effective feature learning within the point cloud representation.

For the \dNdx\ reconstruction, each \gls{tpc} hit is labeled either as positive or negative according to the number of primary electrons it contains, as determined from the MC truth information. Building on previous approaches, we streamline the reconstruction by unifying the two-step procedure described in ref.~\cite{tian2025cluster} into a single model. In this formulation, the reconstruction task can be defined as either a node classification problem or, equivalently, a point cloud segmentation problem.

\subsection{Backbone Structure}
\label{sec:sub-arch}
\begin{figure} [htb]
  \centering
  \includegraphics[width=0.6\textwidth]{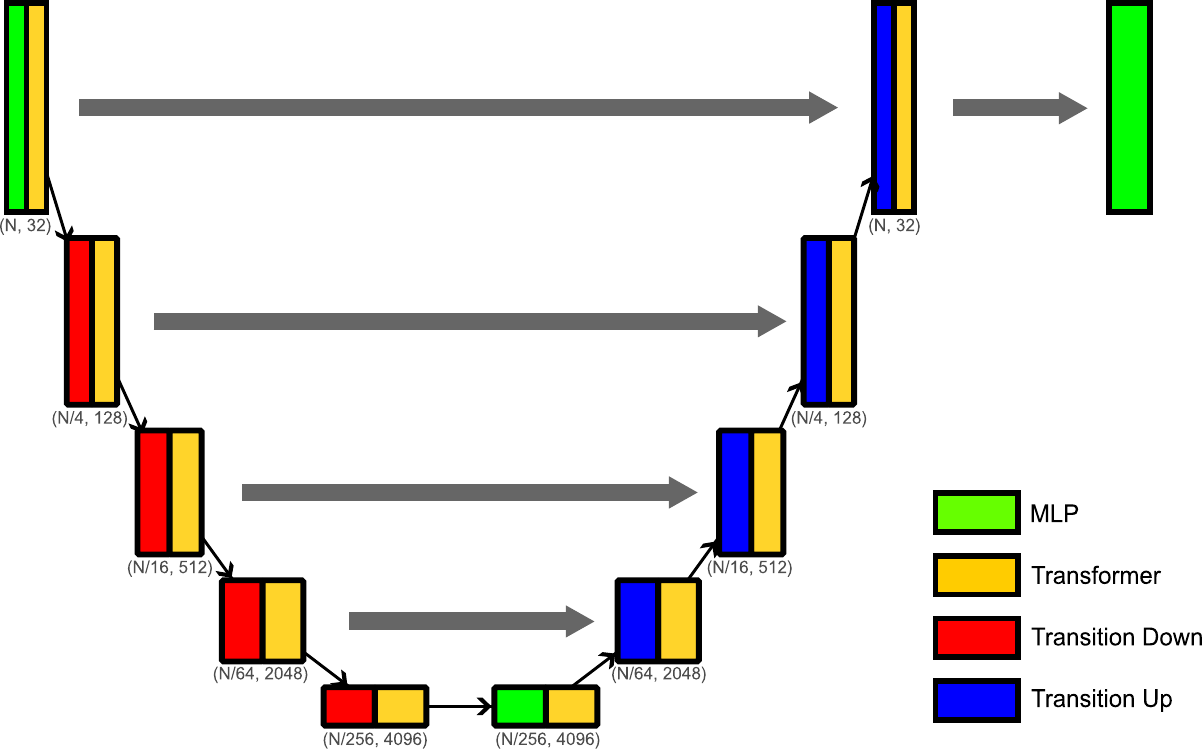}
  \caption{Backbone structure of the \gls{gpt}. The network adopts a U-shaped encoder–decoder architecture with skip connections. Green boxes represent \gls{mlp} layers, yellow boxes indicate transformer layers, red boxes correspond to transition-down layers, and blue boxes correspond to transition-up layers. Within the encoder, nodes are randomly dropped from the graph, and the features are projected into a higher-dimensional space. The decoder performs the inverse operation. Finally, an additional \gls{mlp} layer is appended at the end of the U-Net.}
  \label{fig:backbone}
\end{figure}

As shown in figure~\ref{fig:backbone}, the backbone of the network is a U-Net-based~\cite{ronneberger2015u} hierarchical architecture, consisting of a U-shaped encoder–decoder with skip connections. To begin, a graph is constructed for each \gls{tpc} track, where every \gls{tpc} hit represented as a node. The node features include the hit's charge and timing, and nodes are connected to their $k$ nearest neighbors ($k$NN) based on \gls{3d} Euclidean distance. The graph sequentially executes the hierarchical encoder and decoder. 

Within the encoder, transitions down are performed by applying farthest point sampling to a subset of nodes, while simultaneously projecting their features into a higher-dimension space. Conversely, in the decoder, transitions up are achieved through interpolation, with features mapped back into a lower-dimension space. At each transition, a new $k$NN graph is constructed. Skip connections link layers at corresponding hierarchical levels, ensuring information flow across the network. Additionally, each transition layer incorporates a transformer layer, which will be discussed in the following section. At the end of the decoder, the output graph contains the same number of nodes as the initial input graph. Finally, an \gls{mlp} layer followed by a softmax layer produces probabilities in the range [0, 1]. The detailed network parameters are summarized in table~\ref{tab:model_par}.

\begin{table} [htb]
  \caption{Parameters of \gls{gpt} with transformer layers using subtraction and dot-product operators. In the table, $d_{\textrm{out}}$ denotes the dimension of the output feature.}
  \centerline{
  \begin{tabular}{l|l|c|c}
    \hline
    \multicolumn{2}{c|}{\textbf{Network Parameter}} & \textbf{\gls{gpt} (Sub.)} & \textbf{\gls{gpt} (Dot-prod.)} \\
    \hline
    \multirow{4}{*}{\makecell[l]{Feature\\Dimension}} & Input & 2 & 2 \\
    & Encoder & [32, 128, 512, 2048, 8192] & [8, 32, 128, 512, 2048] \\
    & Decoder & [8192, 2048, 512, 128, 32] & [2048, 512, 128, 32, 8] \\
    & Output & 1 & 1 \\
    \hline
    \multirow{2}{*}{\makecell[l]{Graph\\Construction}} & $k$ in $k$NN graph & 16 & 16 \\
    & Trans. down (up) ratio & 0.25 (4) & 0.25 (4) \\
    \hline
    \multirow{3}{*}{\makecell[l]{Transformer\\Layer}} & Number of heads & \textemdash & 4 \\
    & Attention network $\delta$ & MLP([$d_{\textrm{out}}$, 64, $d_{\textrm{out}}$]) & \textemdash \\
    & Positional network $\theta$ & MLP([3, 64, $d_{\textrm{out}}$]) & MLP([3, 64, 64, $d_{\textrm{out}}$]) \\
    \hline
  \end{tabular}}
  \label{tab:model_par}
\end{table}

\subsection{Transformer Layer}
\label{sec:sub-agg}

The core component of the \gls{gpt} architecture is the transformer layer. In the context of \glspl{gnn}, the transformer layer performs message passing among nodes: each node receives information from its neighbors and aggregates it to update its own representation. The message-passing process can be formalized as:
\begin{equation}
    \mathbf{x_{i}^{'}}=\beta(\mathbf{x_i}) + \sum_{j \in \mathcal{N}(i)}\alpha(\mathbf{x_i}, \mathbf{x_j})\beta(\mathbf{x_j}),
\end{equation}
where $i$ denotes the index of the aggregation node with feature vector $\mathbf{x_i}$ and $\mathcal{N}(i)$ represents its neighboring nodes. The function $\beta$ is an \gls{mlp} that maps the original feature vectors into a latent space, while the function $\alpha$ computes the attention weights used to aggregate neighbor information. The function $\alpha$ is defined as:
\begin{equation}
    \alpha(\mathbf{x_i, x_j})=\textrm{softmax}(\gamma(\mathbf{x_i, x_j})),
\end{equation}
where $\gamma$ is an operator that relates $\mathbf{x_i}$ and $\mathbf{x_j}$. In this paper, two operators are investigated for the \dNdx\ reconstruction.

\paragraph{Subtraction operator.} The first operator is the subtraction operator that is used in \gls{pt}~\cite{zhao2021point}, which is defined as:
\begin{equation}
    \gamma(\mathbf{x_i, x_j})=\delta(\phi(\mathbf{x_i})-\psi(\mathbf{x_j})+\theta(\mathbf{p_i-p_j})),
\end{equation}
where $\phi$ and $\psi$ are neural networks that transform the input features, $\theta$ is a positional embedding network that encodes the relative positional differences between a central node and its neighbors, and $\delta$ is a neural network that integrates these components into a unified representation.

\paragraph{Dot-product operator.} The dot-product operator, which is the fundamental mathematical component underlying self-attention, is further examined. It is defined as:
\begin{equation}
    \gamma(\mathbf{x_i, x_j})=\frac{\phi(\mathbf{x_i})^{\top}(\mathbf{\psi(\mathbf{x_j})+\theta(\mathbf{p_i-p_j})})}{\sqrt{d_{\textrm{out}}}},
\end{equation}
where $d_{\textrm{out}}$ denotes the dimensionality of the output tensor. In addition, the multi-head mechanism is incorporated, executing the attention operation multiple times in parallel, each with its own set of learned projections for queries, keys, and values, and concatenating the outputs of all heads. This design enables the model to capture long-range dependencies and subtle patterns, yielding richer and more expressive representations than the one with the subtraction operator.

\subsection{Training}
\label{sec:sub_train}
In this study, end-to-end training is performed, as perfect labels are available from the full MC simulation. A pad is labeled as positive if it is activated by one or more electrons; otherwise, it is labeled as negative. The neural network outputs a probability for each pad, and binary cross-entropy is employed as the loss function. 

Network training is performed with the AdamW optimizer~\cite{loshchilov2017decoupled}, combined with a decaying learning rate schedule. For the model with the subtraction operator, the learning rate decreases from $10^{-3}$ to $10^{-5}$ over 120 epochs. For the model with the dot-product operator, it decreases from $10^{-4}$ to $10^{-6}$ over 240 epochs.

\section{Performances}
\label{sec:performance}
The performance is evaluated using the testing dataset described in section~\ref{sec:sub-sample}. These samples are generated independently from those used in neural network training. We first optimize the algorithm parameters, then compute the metrics for the binary classification task, and finally present the resulting \gls{pid} performance.

\subsection{Parameter optimization}
\label{sec:sub-op}
\begin{figure} [htb]
  \centering
  \includegraphics[width=0.8\textwidth]{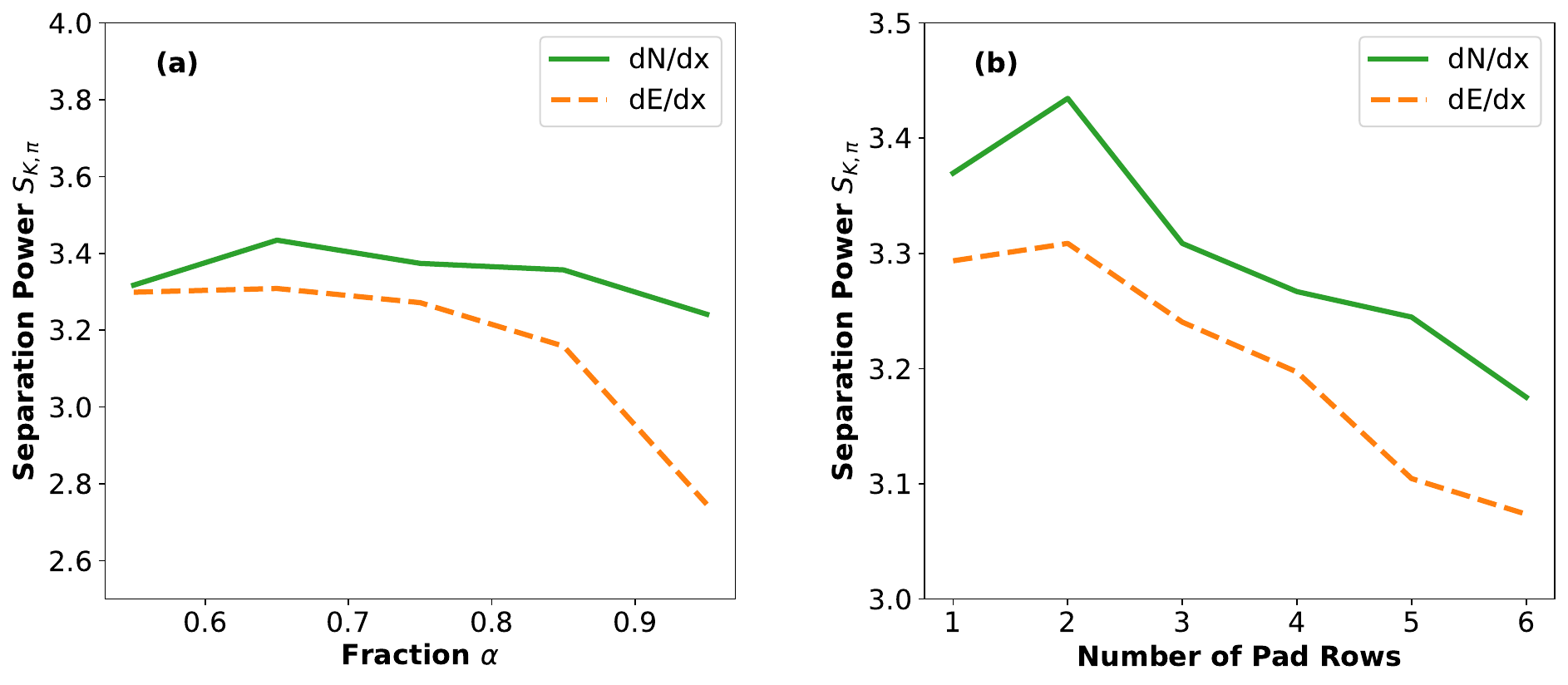}
  \caption{Optimization of the truncated mean method using 20 \gev\ samples. (a) The $K/\pi$ separation power as a function of the fraction $\alpha$, with the number of pad rows fixed to two. (b) The $K/\pi$ separation power as a function of the number of pad rows to be combined, with the fraction fixed at $\alpha = 0.65$. The green solid line represents the \dNdx\ method, while the orange dashed line represents the \dEdx\ method. The optimal values are $\alpha = 0.65$ and two pad rows, respectively. }
  \label{fig:tm_op}
\end{figure}

For the truncated mean method, two free parameters must be optimized. The first is the fraction $\alpha$, typically chosen between 0.50 and 0.85. The second is the number of pad rows combined into a single measurement $M$. In a high-granularity \gls{tpc}, using only one pad row per sample can introduce large fluctuations, whereas combining an appropriate number of pad rows improves stability. In practice, one of these two parameters is optimized while the other is kept fixed, according to the separation power of two particles $A$ and $B$, which is defined as:
\begin{equation}
\label{eq:sp}
    S_{A, B}=\frac{|\mu_{A}-\mu_{B}|}{\sqrt{(\sigma_{A}^{2}+\sigma_{B}^{2})/2}},
\end{equation}
where $\mu_{A (B)}$ and $\sigma_{A (B)}$ are the measured \dEdx\ or \dNdx\ and its resolution for particle $A (B)$, respectively. As shown in figure \ref{fig:tm_op}, the optimal fraction $\alpha$ is 0.65, with two pad rows combined.

\begin{figure} [htp]
  \centering
  \includegraphics[width=0.6\linewidth]{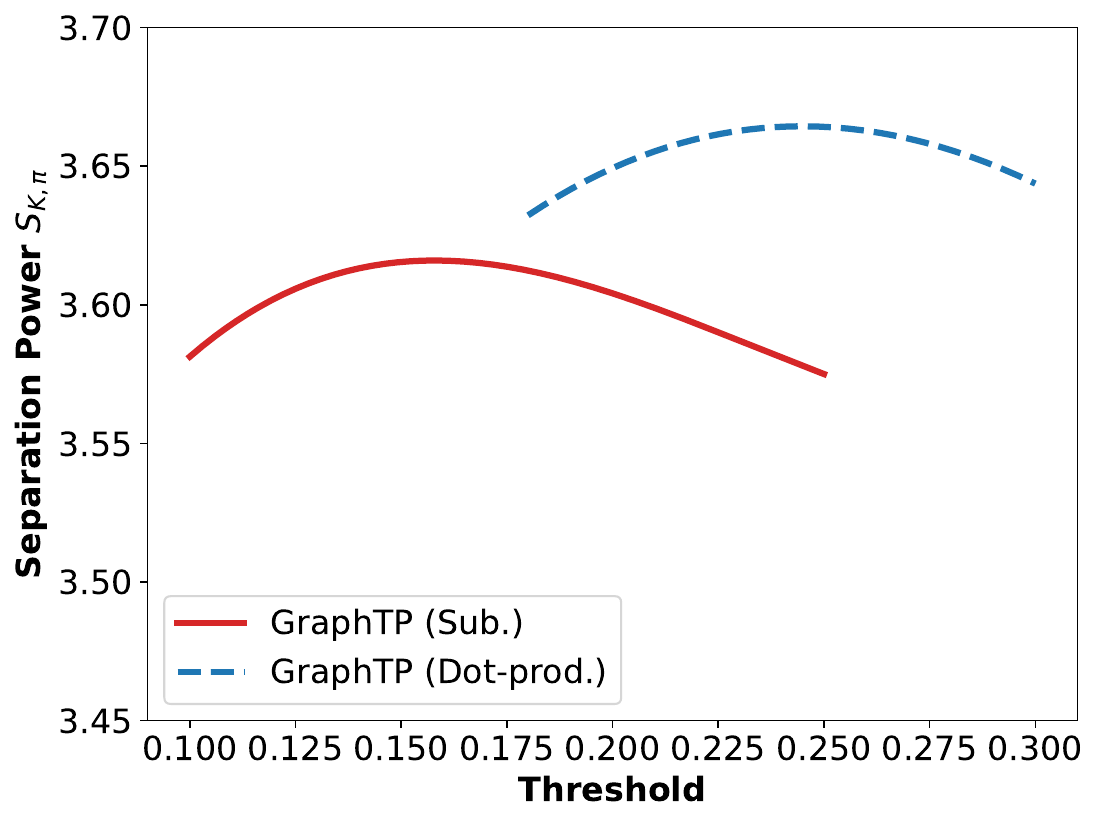}
  \caption{Threshold optimization in \gls{gpt} using 20 \gev\ samples. The red solid line shows the \gls{gpt} using the subtraction operator, while the blue dashed line corresponds to the \gls{gpt} with the dot-product operator. The respective optimal thresholds are 0.157 and 0.261.}
  \label{fig:ml_op}
\end{figure}

For the \gls{gpt}, the neural network outputs a probability $p_i$ for each \gls{tpc} hit. These probabilities are then used for the \dNdx\ reconstruction, which is defined as:
\begin{equation}
    \text{d}N/\text{d}x = \frac{1}{L}\sum_{i=1}^{N} \mathbf{1}_{\{p_i > \tau\}},
\end{equation}
where $i$ denotes the hit index, $\mathbf{1}$ is the indicator function, $\tau$ is a threshold, and $L$ is the track length. The threshold $\tau$ must be optimized according to the separation power. Figure~\ref{fig:ml_op} illustrates the $K/\pi$ separation power as a function of the threshold. The optimal values are 0.157 and 0.261 for the \gls{gpt} with the subtraction operator and dot-product operator, respectively.

\subsection{Classification Performance}
\label{sec:cls-perf}

\begin{figure} [htb]
    \centering
    \includegraphics[width=0.75\linewidth]{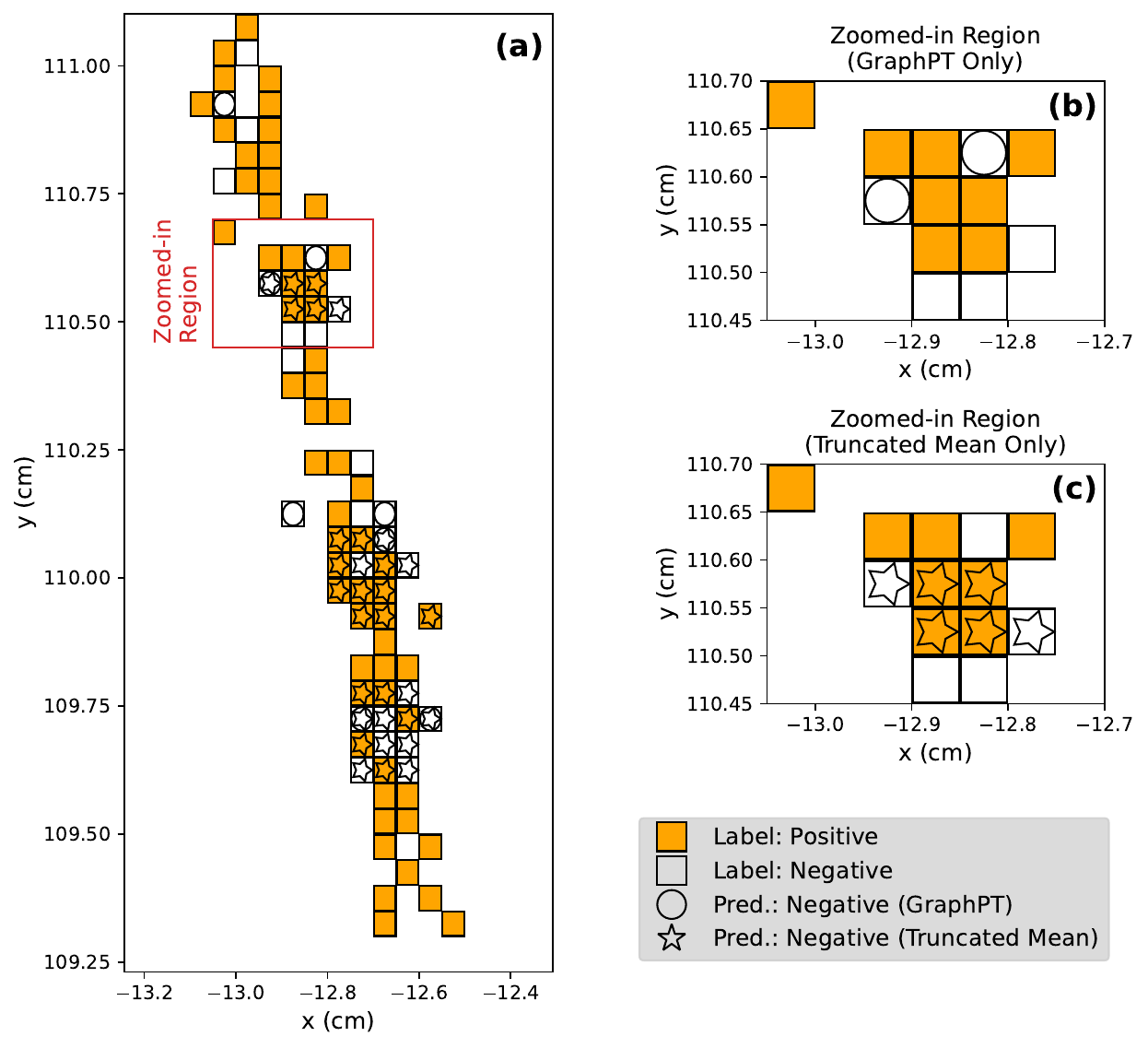}
    \caption{Predicted pad classes for the same track segment shown in figure~\ref{fig:tpc_readout}. (a) Pads predicted as negatives are marked by circles for the \gls{gpt} and stars for the truncated mean method, respectively; all other pads are predicted as positives. The red box highlights the zoomed-in region. (b) The zoomed-in region using only the \gls{gpt} method. (c) The zoomed-in region using only the truncated mean method.}
    \label{fig:readout_rec}
\end{figure}

With the optimized parameters, \dNdx\ reconstruction is carried out using both the truncated mean method and the \gls{gpt} model. Figure~\ref{fig:readout_rec} shows an example of the classification result for the track segment shown in figure~\ref{fig:tpc_readout}: pads marked as circles (for the \gls{gpt}) or stars (for the truncated mean) are classified as negatives (secondary electrons), while all other pads are classified as positives (primary electrons). Examination of the zoomed-in region in figure~\ref{fig:readout_rec}b reveals that the \gls{gpt} model produces almost no false-negatives, whereas the truncated mean method is more aggressive and results in several false-negatives, as illustrated in figure~\ref{fig:readout_rec}c, which can reduce signal efficiency.

The classification performance is further validated on the full testing samples using the two widely adopted evaluation metrics: accuracy and the F1-score. Accuracy measures the proportion of correct predictions out of all predictions, making it a intuitive and widely applicable metric. It is defined as:
\begin{equation}
    \textrm{ACC}=\frac{\textrm{TP + TN}}{\textrm{TP + TN + FP + FN}},
\end{equation}
where the TP, TN, FP, FN denote the numbers of true positives, true negatives, false positives and false negatives, respectively. The F1-score, by contrast, is the harmonic mean of precision and recall, capturing the balance between correctly identifying positive cases and avoiding false alarms:
\begin{equation}
    \textrm{F1-Score}=2\times\frac{P \times R}{P+R},
\end{equation}
where $P$ denotes precision ($\textrm{TP / (TP + FP)}$) and $R$ denotes recall ($\textrm{TP / (TP + FN)}$). 

Table~\ref{tab:cls-metrics} presents the results obtained with different reconstruction methods. Relative to the truncated mean approach, the \gls{gpt} models yield lower precision but substantially higher recall, indicating markedly improved signal efficiency at the cost of slightly reduced purity. This observation is consistent with the behavior illustrated in figure~\ref{fig:readout_rec}. In terms of overall classification performance—assessed by accuracy and F1-score—the \gls{gpt} models clearly surpass the traditional truncated mean method. Among the two \gls{gpt} variants, the model incorporating the dot-product operator achieves slightly superior classification performance.

\begin{table} [htb]
  \caption{Classification metrics.}
  \centering
  \begin{tabular}{l|c|c|c|c}
    \hline
    \textbf{Method} & \textbf{Accuracy} & \textbf{Precision} & \textbf{Recall} & \textbf{F1-Score} \\
    \hline
    Truncated Mean & 0.601 & 0.743 & 0.574 & 0.648 \\
    \gls{gpt} (Sub.) & 0.698 & 0.689 & 0.960 & 0.802 \\
    \textbf{\gls{gpt} (Dot-prod.)} & \textbf{0.707} & \textbf{0.702} & \textbf{0.941} & \textbf{0.804} \\
    \hline
  \end{tabular}
  \label{tab:cls-metrics}
\end{table}

\subsection{PID Performance}
\label{sec:pid_perf}
\begin{figure} [htb]
  \centering
  \includegraphics[width=0.8\textwidth]{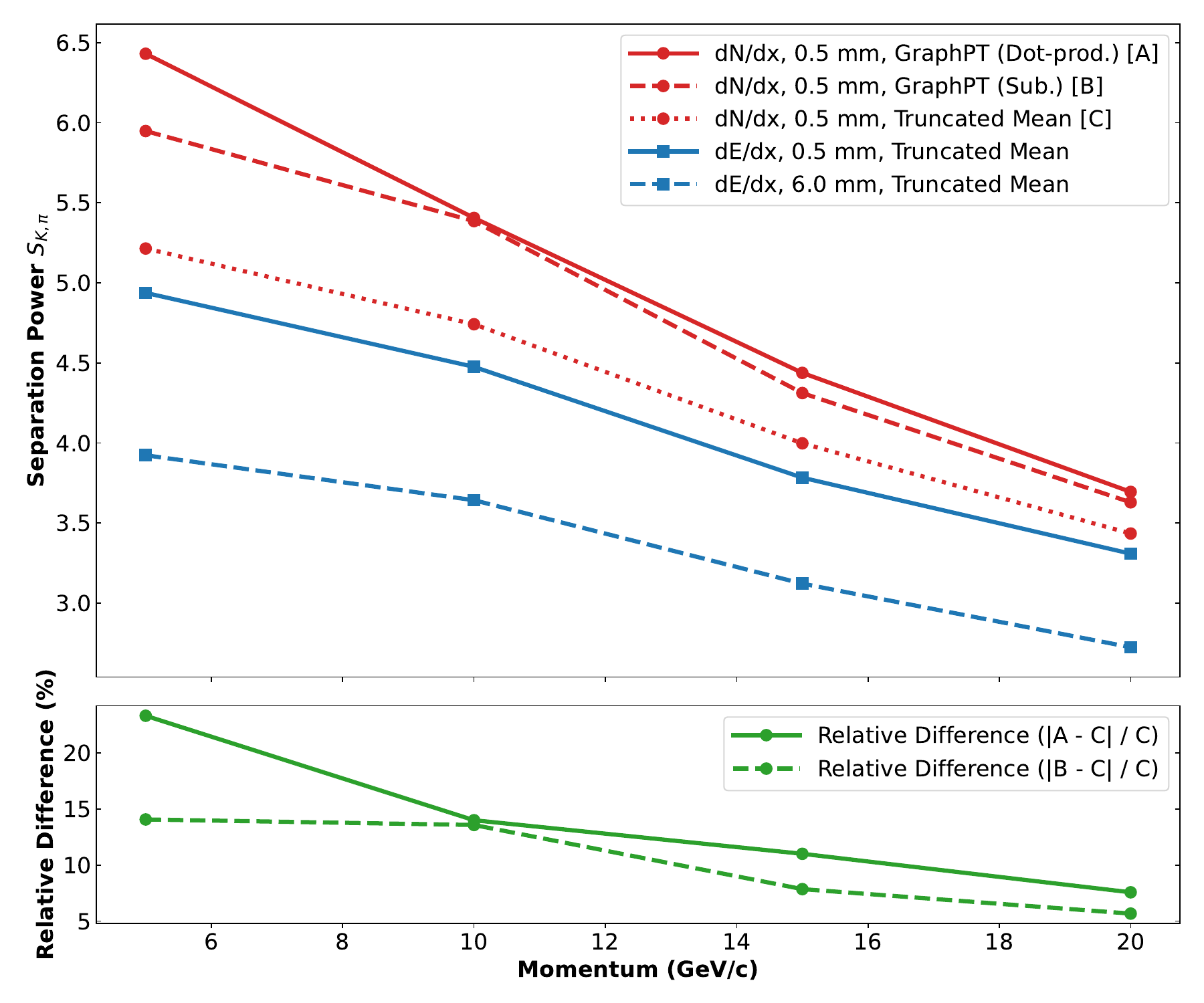}
  \caption{$K/\pi$ separation power as a function of particle momentum. Red lines with circular markers denote the \dNdx\ results, while blue lines with rectangular markers correspond to the \dEdx\ results. For the red curves, the solid line shows the \gls{gpt} model with a subtraction operator, the dashed line represents the \gls{gpt} model with a dot-product operator, and the dotted line indicates the traditional truncated mean method. For the blue curves, the solid and dashed lines correspond to \dEdx\ results obtained with the truncated mean method for pad lengths of 500~\mum\ and 6~mm, respectively. The relative differences between the \gls{gpt} model with a dot-product (subtraction) operator and the truncated mean \dNdx\ are shown by the solid (dashed) green line.}
  \label{fig:sp}
\end{figure}

The \gls{pid} performance is further evaluated using the particle separation power defined in eq.~\ref{eq:sp}. Figure~\ref{fig:sp} presents the $K/\pi$ separation power $S_{K,\,\pi}$ as a function of track momentum. The tracks originate from the collision point with a polar angle of $60^{\circ}$, corresponding to a track length of approximately 1.4~m. For the \dNdx\ results (shown in red circles), the conclusions are consistent with those in table~\ref{tab:cls-metrics}: the \gls{gpt} method significantly enhances \gls{pid} performance compared to the traditional truncated mean approach, with the dot-product operator yielding the best overall $K/\pi$ separation. Specifically, the improvements range from approximately 10\% (5\%) to 20\% (15\%) in the momentum interval from 5 to 20~\gev\ for the dot-product (subtraction) operator.

For comparison, \dEdx\ results obtained with pad sizes of $500 \times 500$~\mumsq\ and $6 \times 1\,\textrm{mm}^2$ are also shown (shown in blue rectangles), with the latter representing a conventional large-pad readout. As expected, all \dNdx\ results outperform those from \dEdx. Moreover, the \gls{pid} performance achieved with \dNdx\ is even superior when compared to \dEdx\ obtained using the conventional large-pad readout. 

To further assess the robustness of the \gls{gpt} model, we apply it  to a finer-granularity readout with a pad size of $200 \times 200$~\mumsq. An even larger performance gain is observed with the \gls{gpt} model. Additional details are provided in appendix~\ref{sec:appendix}.

\section{Conclusion and Outlook}
\label{sec:summary}
In this paper, we propose a general deep-learning–based \dNdx\ reconstruction approach, \gls{gpt}, for high-granularity \glspl{tpc}. The \gls{tpc} data are represented as point clouds. The network backbone employs a U-Net architecture built upon \glspl{gnn}, incorporating an attention mechanism for node aggregation specifically tailored to point cloud processing. The model is optimized for the \gls{cepc} \gls{tpc} and trained on hadron samples generated with a sophisticated simulation framework that provides detailed detector simulation and digitization.

By evaluating both the classification and \gls{pid} performances, we find that the \gls{gpt} model achieves superior results. Compared with the traditional truncated mean method, the \gls{gpt} model serves as a substantially more powerful classifier, delivering higher accuracy and F1-score. The $K/\pi$ separation power improves by approximately 10\% (5\%) to 20\% (15\%) in the momentum interval from 5 to 20~\gev\ when using dot-product (subtraction) operators. Relative to the conventional \dEdx\ method for the current-generation detector, the improvement is even larger, reaching nearly 50\%. 

Looking ahead, several aspects require further study. First, the \gls{gpt} model itself still needs optimization in terms of network architecture and hyperparameters. Second, in this work the model was trained on a fix-angle sample; future studies should extend the training to much larger and more diverse datasets. This will also necessitate improving the computational efficiency of the model to enable large-scale training. Moreover, as with any supervised-learning approach, the performance of the model depends on the consistency between the training data and the data it is applied to. Periodic retraining and calibration with real data will be essential to maintain robustness against systematic effects such as spacial inhomogeneities in detector response, time variations in TPC operating conditions, and field distortions induced by ion backflow.

The ultimate goal of this study is to access the feasibility of \gls{pid} performance using the \dNdx\ method. To this end, further validation against real detectors is essential. For the \gls{cepc} \gls{tpc}, two TPC prototypes have already been designed. Beam tests are planned to evaluate their performance and provide experimental validation of the high-granularity readout approach. One such beam test has scheduled at Deutsches Elektronen-Synchrotron (DESY)~\cite{diener2019desy}, within the LCTPC Collaboration and the CERN R\&D Collaboration on gaseous detector (DRD1)~\cite{colaleo2024drd1}.

\acknowledgments
This research was funded by National Natural Science Foundation of China (NSFC) under Contract Nos. 12475200 and Nos. 12475192. We sincerely thank Tianchi Zhao for the helpful discussions and his valuable suggestions.

\appendix
\section{Validation Using a Reduced Pad Size of $200 \times 200$~\mumsq}
\label{sec:appendix}

\begin{figure} [htb]
    \centering
    \includegraphics[width=0.75\linewidth]{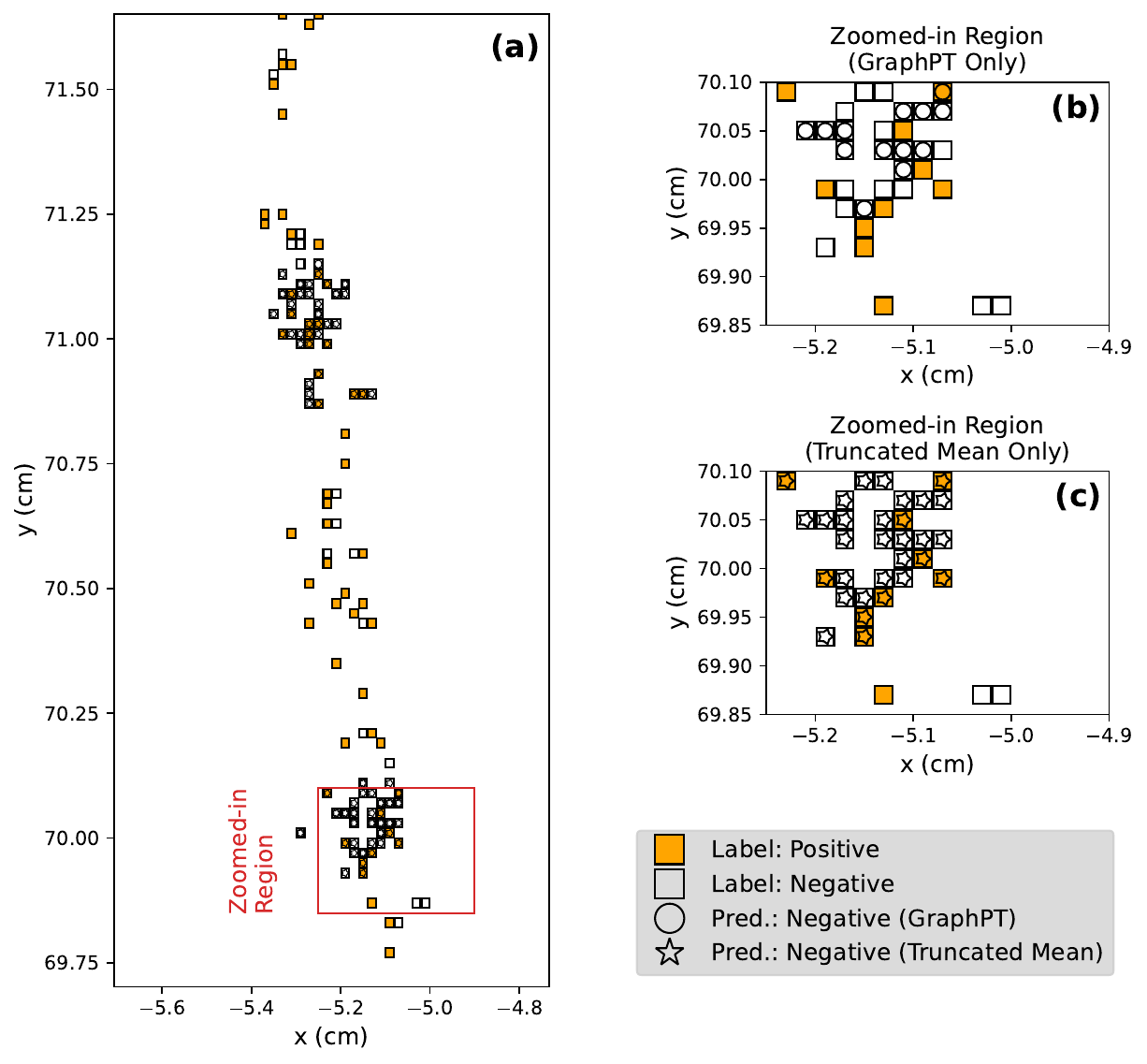}
    \caption{Predicted pad classes for a track segment with a pad size of $200 \times 200$~\mumsq. (a) Pads predicted as negatives are marked by circles for the \gls{gpt} and stars for the truncated mean method, respectively; all other pads are predicted as positives. The red box highlights the zoomed-in region. (b) The zoomed-in region using only the \gls{gpt} method. (c) The zoomed-in region using only the truncated mean method.}
    \label{fig:readout_rec_200um}
\end{figure}

To further validate the \gls{gpt} model, an alternative simulation configuration is employed. In this setup, the detector is simulated with a smaller pad size of $200 \times 200$~\mumsq. As shown in figure~\ref{fig:readout_rec_200um}, the increased granularity retains more of the underlying raw information: the hits become sparser, and both primary and secondary electrons are more readily detected. This richer but less compressed input makes the reconstruction task more challenging. In this simulation, only the pad size is modified; all other aspects of the detector and electronics design remain unchanged. Therefore, aside from the reduced pad size, all parameters are identical to those used in the default simulation. 

The neural network architecture follows the design described in section~\ref{sec:sub-arch}. For the self‑attention block, only the dot‑product operator—--shown to deliver state‑of‑the‑art performance—--is used. The model is trained with a learning rate decreasing from $10^{-4}$ to $10^{-5}$ over 140 epochs. Following the procedure described in section~\ref{sec:sub-op}, the threshold $\tau$ is optimized to $10^{-4}$; the fraction $\alpha$ and the number of combined pad rows used in the truncated mean method are optimized to 0.60 and 4, respectively. 

\begin{table} [htb]
  \caption{$K/\pi$ separation power obtained using the \dNdx\ method with the truncated mean (TM) and the GraphPT (ML) model, along with their relative difference, defined as |ML - TM|/TM.}
  \centering
  \begin{tabular}{l|c|c|c|c} 
  \hline 
  \multirow{2}{*}{\makecell[l]{\textbf{Pad Length = 500 \mum}}}
  & \multicolumn{4}{c}{\textbf{Momentum (GeV/c)}} \\ 
  \cline{2-5} 
  & \textbf{5} & \textbf{10} & \textbf{15} & \textbf{20} \\ 
  \hline 
  Truncated Mean & $5.21 \pm 0.04$ & $4.74 \pm 0.04$ & $4.00 \pm 0.04$ & $3.43 \pm 0.03$ \\ 
  \gls{gpt} (Dot-prod.) & $6.43 \pm 0.05$ & $5.40 \pm 0.04$ & $4.44 \pm 0.04$ & $3.69 \pm 0.03$ \\ 
  Relative Difference (\%) & $23.3 \pm 1.4$ & $14.0 \pm 1.3$ & $11.0 \pm 1.4$ & $7.6 \pm 1.4$ \\ 
  \hline 
  \multirow{2}{*}{\makecell[l]{\textbf{Pad Length = 200 \mum}}}
  & \multicolumn{4}{c}{\textbf{Momentum (GeV/c)}} \\ 
  \cline{2-5} 
  & \textbf{5} & \textbf{10} & \textbf{15} & \textbf{20} \\ 
  \hline 
  Truncated Mean & $5.04 \pm 0.04$ & $4.60 \pm 0.04$ & $3.93 \pm 0.04$ & $3.42 \pm 0.04$ \\ 
  \gls{gpt} (Dot-prod.) & $6.83 \pm 0.05$ & $5.90 \pm 0.05$ & $4.79 \pm 0.04$ & $3.94 \pm 0.03$ \\ 
  Relative Difference (\%) & $35.5 \pm 1.5$ & $28.3 \pm 1.5$ & $21.7 \pm 1.5$ & $15.2 \pm 1.5$ \\
  \hline 
  \end{tabular}
  \label{tab:sp-200um}
\end{table}

The classification results for a track segment are shown in figure~\ref{fig:readout_rec_200um}. A conclusion similar to the $500 \times 500$~\mumsq\ case can be drawn: the \gls{gpt} model produces almost no false-negatives, whereas the truncated mean method is more aggressive and yields several false-negative classifications, thereby reducing signal efficiency. Table~\ref{tab:sp-200um} summarizes the resulting $K/\pi$ separation power. Compared with the $500 \times 500$~\mumsq\ configuration, the truncated mean performance is slightly degraded. This is mainly due to the reduced signal-to-noise ratio: although the noise level per pad remains $100~e^-$, the total number of pads increases by a factor of 6.25. Because the truncated mean method suppress secondary-electron contributions by removing entire row pads, it cannot fully exploit the high-granularity information. In contrast, the \gls{gpt} model can perform pad-by-pad suppression, leading to a significantly improved $K/\pi$ separation power for the $200 \times 200$~\mumsq\ readout. The improvement is approximately 35\% to 15\% across the momentum range from 5 to 20~\gev.

\paragraph{Code Availability Statement} This article has associted code in a code repository. The code for the \gls{gpt} model is publicly available at \url{https://github.com/littlepi/TpcDndxGraphPT}.


\bibliographystyle{JHEP}
\bibliography{biblio.bib}

\end{document}